\documentclass[12pt]{article}
\usepackage{amssymb}
\begin{document}
\thispagestyle{empty}
\newcommand{\be}{\begin{equation}}
\newcommand{\ee}{\end{equation}}
\newcommand{\sect}[1]{\setcounter{equation}{0}\section{#1}}
\newcommand{\vs}[1]{\rule[- #1 mm]{0mm}{#1 mm}}
\newcommand{\hs}[1]{\hspace{#1mm}}
\newcommand{\mb}[1]{\hs{5}\mbox{#1}\hs{5}}
\newcommand{\bea}{\begin{eqnarray}}
\newcommand{\eea}{\end{eqnarray}}
\newcommand{\wt}[1]{\widetilde{#1}}
\newcommand{\ux}[1]{\underline{#1}}
\newcommand{\ov}[1]{\overline{#1}}
\newcommand{\sm}[2]{\frac{\mbox{\footnotesize #1}\vs{-2}}
           {\vs{-2}\mbox{\footnotesize #2}}}
\newcommand{\prt}{\partial}
\newcommand{\eps}{\epsilon}\newcommand{\p}[1]{(\ref{#1})}
\newcommand{\R}{\mbox{\rule{0.2mm}{2.8mm}\hspace{-1.5mm} R}}
\newcommand{\Z}{Z\hspace{-2mm}Z}
\newcommand{\cd}{{\cal D}}
\newcommand{\cg}{{\cal G}}
\newcommand{\ck}{{\cal K}}
\newcommand{\cw}{{\cal W}}
\newcommand{\vj}{\vec{J}}
\newcommand{\vl}{\vec{\lambda}}
\newcommand{\vz}{\vec{\sigma}}
\newcommand{\vt}{\vec{\tau}}
\newcommand{\poiss}{\stackrel{\otimes}{,}}
\newcommand{\tx}{\theta_{12}}
\newcommand{\tb}{\overline{\theta}_{12}}
\newcommand{\zw}{{1\over z_{12}}}
\newcommand{\sqp}{{(1 + i\sqrt{3})\over 2}}
\newcommand{\sqm}{{(1 - i\sqrt{3})\over 2}}
\newcommand{\NP}[1]{Nucl.\ Phys.\ {\bf #1}}
\newcommand{\PLB}[1]{Phys.\ Lett.\ {B \bf #1}}
\newcommand{\PLA}[1]{Phys.\ Lett.\ {A \bf #1}}
\newcommand{\NC}[1]{Nuovo Cimento {\bf #1}}
\newcommand{\CMP}[1]{Commun.\ Math.\ Phys.\ {\bf #1}}
\newcommand{\PR}[1]{Phys.\ Rev.\ {\bf #1}}
\newcommand{\PRL}[1]{Phys.\ Rev.\ Lett.\ {\bf #1}}
\newcommand{\MPL}[1]{Mod.\ Phys.\ Lett.\ {\bf #1}}
\newcommand{\BLMS}[1]{Bull.\ London Math.\ Soc.\ {\bf #1}}
\newcommand{\IJMP}[1]{Int.\ J.\ Mod.\ Phys.\ {\bf #1}}
\newcommand{\JMP}[1]{Jour.\ Math.\ Phys.\ {\bf #1}}
\newcommand{\LMP}[1]{Lett.\ Math.\ Phys.\ {\bf #1}}
\newpage
\setcounter{page}{0} \pagestyle{empty} \vs{12}
\begin{center}
{\LARGE {\bf \quad\quad Quaternionic and Octonionic
Spinors.\quad\quad A Classification}}\\ {\quad}\\

\vs{10} {\large H.L. Carrion, M. Rojas and F. Toppan} ~\\ \quad
\\
 {\large{\em CBPF - CCP}}\\{\em Rua Dr. Xavier Sigaud
150, cep 22290-180 Rio de Janeiro (RJ)}\\{\em Brazil}\\

\end{center}
{\quad}\\
\centerline{ {\bf Abstract}}

\vs{6}

Quaternionic and octonionic realizations of Clifford algebras and
spinors are classified and explicitly constructed in terms of
recursive formulas. The most general free dynamics in arbitrary
signature space-times for both quaternionic and octonionic spinors
is presented. In the octonionic case we further provide a
systematic list of results and tables expressing, e.g., the
relations of the octonionic Clifford algebras with
the $G_2$ cosets over the Lorentz algebras, the
identities satisfied by the higher-rank antisymmetric octonionic
tensors and so on. Applications of these results range from the
classification of octonionic generalized supersymmetries, the
construction of octonionic superstrings, as well as the
investigations concerning the recently discovered octonionic
$M$-superalgebra and its superconformal extension. \vs{6} \vfill
\rightline{CBPF-NF-002/03} {\em E-mails:}{ hleny@cbpf.br,
mrojas@cbpf.br, toppan@cbpf.br}
\pagestyle{plain}
\renewcommand{\thefootnote}{\arabic{footnote}}

\section{Introduction.}

The unification program aiming at a unified description of the
known interactions as well as a consistent quantum formulation for
gravity, nowadays mostly points towards higher-dimensional
supersymmetric theories. At present the most promising, however
still conjectural, candidate should live in eleven dimensions and
goes under the name of $M$-theory \cite{bv}. The theoretical (and
phenomenological) consistency requirements put on any possible
candidate for unification necessarily lead to a systematic
investigation of the properties of Clifford algebras and spinors
in space-times of arbitrary dimension and signature. Exploring in
full generality the existence of specific algebraic relations
(such as the identities necessary to prove the $k$-symmetry
invariance in the GS formulation of the superstring, see
\cite{gsw}), which are technically relevant in the model
construction, is a necessary preliminary mathematical step before
any attempt to model building.\par From a mathematical point of
view, Clifford algebras have been classified in the sixties (see
\cite{abs}). Some twenty years later the relation between
supersymmetry and division algebras was analyzed in \cite{kt}. A
systematic and very convenient presentation in physicists'
notation of the classification for Clifford algebras and spinors,
based on the three associative division algebras of the real,
complex and quaternionic numbers (${\bf R}$, ${\bf C}$ and ${\bf
H}$), can be found in \cite{oku1}. This relatively old subject has
been revived recently in a series of work (\cite{fer}). The aim in
this case was the classification (once again based on the ${\bf
R}$, ${\bf C}$, ${\bf H}$ division algebras) of the generalized
supersymmetries admitting the presence of tensorial bosonic
central charges and going therefore beyond the standard H\L S
scheme \cite{hls}. The real-valued $M$-algebra underlining the
$M$-theory is the most celebrated and possibly the most physically
relevant example in this class of generalized supersymmetries. In
the last few months it was pointed out in \cite{lt1} and
\cite{lt2} that the $M$-algebra admits a consistent octonionic
restriction with surprising properties, which will be discussed in
the following.\par The first attempt of introducing octonions in
physics goes back to the works of Jordan \cite{jor}. More
recently, and in connection with the specific program of
unification through supersymmetry, we can cite a series of works
\cite{{fm},{cs}} devoted to the octonionic description of the
superstring. Already in \cite{kt} some mathematical results
concerning the relations of the octonions with the Lorentz and
Jordan algebras are mentioned, while a more developed
investigation of this topic is presented in \cite{oku2}. Moreover,
in several different works (see e.g. \cite{{hl},{cp}}) the octonionic
characterization of the seven sphere $S^7$ (regarded as a
compactification space for the eleven-dimensional maximal
supergravity) and the analysis of its properties were
investigated.\par Octonions are non-associative and can not be
represented through matrices with the standard matrix product.
Octonionic realizations of Clifford algebras have peculiar
properties, the most noticeable perhaps is the fact that they do
not generate the corresponding Lorentz group, but only its coset
over $G_2$, the group of automorphisms of the octonions \cite{hl}.
\par This
work is devoted to a systematic investigation of the properties of
the quaternionic and the octonionic realizations of the Clifford
algebras. More specifically, in the first part we classify such
realizations, also furnishing recursive algorithms to explicitly
construct them. Later, quaternionic and octonionic spinors are
introduced. The notion of Weyl projection, whenever applicable,
for these two classes of spinors is defined. The consistency
conditions for the existence of a free dynamics for quaternionic
and octonionic spinors are fully investigated and classified. We
produce a whole set of tables expressing the allowed space-times
admitting kinetic or pseudokinetic, as well as massive or
pseudomassive, terms in the free-spinors lagrangian. These results
can be considered as quaternionic and octonionic extensions of
previous classification schemes available for real-valued spinors,
see \cite{dt}. Since quaternionic Clifford algebras and spinors
can always be represented through real-valued matrices and column
vectors, the tables presented in the quaternionic case can be
recovered from suitably constraining the real-valued case in order
to admit a quaternionic structure. The situation however is
entirely different, for the motivations that we already recalled,
in the octonionic case. Furthermore, we elucidate the connection
of the octonionic realizations of Clifford algebras with the $G_2$
cosets of the Lorentz groups. We also produce highly non-trivial
tables expressing identities for higher rank antisymmetric
octonionic tensors. Some of these identities already found
application in the investigations concerning the octonionic
generalized supersymmetries. As a particular example we can
mention that, in the already recalled octonionic $M$-theory, the
octonionic $5$-brane sector is identified with the octonionic $M1$
and $M2$ sectors. \par The classification of the consistency
conditions for the free octonionic dynamics should be regarded as
a first preliminary step towards the investigation of octonionic
supersymmetric dynamical systems associated to the generalized
octonionic supersymmetries. It is worth stressing the fact that,
for what concerns the latter, for the time being just examples of
such superalgebras, the ones seemingly more attractive on physical
grounds, have been analyzed so far. A classification scheme is
still in progress.\par The paper is organized as follows. In the
next section we review \cite{oku1} the classification of Clifford
algebras and spinors in terms of the associative division
algebras. In section {\bf 3} we present a systematic construction
of the irreducible representations for real-valued Clifford
algebras. This paves us the way to introduce in section {\bf 4}
the explicit construction of the associative quaternionic and the
non-associative octonionic realizations of the Clifford algebras.
In section {\bf 5} we introduce the necessary conventions to
introduce the dynamics for real, quaternionic and octonionic
spinors. In section {\bf 6} the results of \cite{dt} concerning
the classification of the most general free dynamics for real
spinors in arbitrary signature space-times are reviewed. In
section {\bf 7} and {\bf 8} these results are extended to,
respectively, quaternionic and octonionic spinors. Section {\bf 9}
is devoted to present a list of identities, due to the
non-associativity of the octonions, involving higher-rank
antisymmetric octonionic tensors. In the next section ({\bf 10})
some applications of these last results to octonionic generalized
supersymmetries and $M$-theory are mentioned. Finally, in the
Conclusions, we point out possible future developments of the line
of investigation here presented.

\section{On Clifford algebras and division algebras.}

For later convenience we review in this section, following
\cite{oku1}, the classification of the Clifford algebras associated
to the ${\bf R}, ${\bf C}, ${\bf H}$ associative division
algebras.\par The most general irreducible {\em real} matrix
representations of the Clifford algebra
\begin{eqnarray}
\Gamma^\mu\Gamma^\nu+\Gamma^\mu\Gamma^\nu &=& 2\eta^{\mu\nu},
\end{eqnarray}
with $\eta^{\mu\nu}$ being a diagonal matrix of $(p,q)$ signature
(i.e. $p$ positive, $+1$, and $q$ negative, $-1$, diagonal
entries), can be classified according to the property of the most
general $S$ matrix commuting with all the $\Gamma$'s ($\relax
[S,\Gamma^\mu ] =0$ for all $\mu$). If the most general $S$ is a
multiple of the identity, we get the normal (${\bf R}$) case.
Otherwise, $S$ can be the sum of two matrices, the second one
multiple of the square root of $-1$ (this is the almost complex,
${\bf C}$ case) or the linear combination of $4$ matrices closing
the quaternionic algebra (this is the ${\bf H}$ case). According
to \cite{oku1} the {\em real} irreducible representations are of
${\bf R}$, ${\bf C}$, ${\bf H}$ type, according to the following
table, whose entries represent the values $p-q~ mod~ 8$
 { {{\begin{eqnarray}&
\begin{tabular}{|c|c|c|}\hline
${\bf R} $&${\bf C}$ &${\bf H}$\\ \hline $0,2$&$$&$4,6$\\ \hline
$1$&$3,7$&$5$\\ \hline
\end{tabular}&\nonumber\end{eqnarray}}} }

The real irreducible representation is always unique unless
$p-q~mod~8 = 1,5$. In these signatures two inequivalent real
representations are present, the second one recovered by flipping
the sign of all $\Gamma$'s ($\Gamma^\mu \mapsto - \Gamma^\mu$).

Furthermore, in the given signatures $p-q~mod~8 = 0,4,6,7$,
without loss of generality, the $\Gamma^\mu$ matrices can be
chosen block-antidiagonal (Weyl-type matrices), i.e. of the form
\begin{eqnarray}
\Gamma^\mu &=&\left( \begin{array}{cc}
  0 & \sigma^\mu \\
  {\tilde\sigma}^\mu & 0
\end{array}\right)\label{weyl}
\end{eqnarray}
In these signatures it is therefore possible to introduce the
Weyl-projected spinors, whose number of components is half of the
size of the corresponding $\Gamma$-matrices.\par The division
algebra characteristic for spinors (of ${\bf R}$, ${\bf C}$, ${\bf
H}$ type) can be found in \cite{fer}.\par It is useful to
illustrate our discussion presenting a table with the division
algebra characteristic and number of real components for both
Clifford algebras (${\bf \Gamma}$) and fundamental spinors (${\bf
\Psi}$), at least in the specific case of the Minkowskian
spacetimes up to $11$ dimensions. We obtain the following table
 { {{\begin{eqnarray}&
\begin{tabular}{|c|c|c||c|c|c|}\hline
$(p,q) $&${\bf \Gamma}$ &${\bf \Psi}$ &$(p,q)$&${\bf
\Gamma}$&${\bf \Psi}$
\\ \hline $(1,0)$&${\bf R}, 1
$&${\bf R}, 1$&$(0,1)$&${\bf C}, 2 $&${\bf R}, 1$

\\ \hline $(1,1)$&${\bf R}, 2$&${\bf R},1$&$(1,1)$&${\bf R}, 2
$&${\bf R}, 1$\\ \hline

 $(1,2)$&${\bf C}, 4
$&${\bf R}, 2$&$(2,1)$&${\bf R}, 2 $&${\bf R}, 2$\\ \hline

$(1,3)$&${\bf H}, 8$&${\bf C},4$&$(3,1)$&${\bf R}, 4 $&${\bf C},
4$\\ \hline

$(1,4)$&${\bf H}, 8 $&${\bf H}, 8$&$(4,1)$&${\bf C}, 8 $&${\bf H},
8$\\ \hline

$(1,5)$&${\bf H}, 16$&${\bf H},8$&$(5,1)$&${\bf H}, 16 $&${\bf H},
8$\\ \hline

 $(1,6)$&${\bf C}, 16
$&${\bf H}, 16$&$(6,1)$&${\bf H}, 16 $&${\bf H}, 16$\\ \hline

$(1,7)$&${\bf R}, 16$&${\bf C},16$&$(7,1)$&${\bf H}, 32 $&${\bf
C}, 16$\\ \hline

 $(1,8)$&${\bf R}, 16
$&${\bf R}, 16$&$(8,1)$&${\bf C}, 32 $&${\bf R}, 16$\\ \hline

$(1,9)$&${\bf R}, 32$&${\bf R},16$&$(9,1)$&${\bf R}, 32$&${\bf R},
16$\\ \hline

 $(1,10)$&${\bf C}, 64
$&${\bf R}, 32$&$(10,1)$&${\bf R}, 32 $&${\bf R}, 32$\\ \hline

\end{tabular}&\nonumber\end{eqnarray}}} }
It should be noticed that, as far as Clifford algebras are
concerned, the above table is not symmetric under the exchange
$(p,q)\leftrightarrow (q,p)$ (the simplest example is the
one-dimensional Clifford algebra with negative eigenvalue,
represented by a $2\times 2$ real matrix). On the other hand, the
properties of spinors are invariant (in some of the cases, for the
signatures allowing it, the Weyl projection is required). As a
consequence, the theories under consideration can be equivalently
described either working with the $(p,q)$ or with the $(q,p)$
signatures.\par For what concerns the generalized supersymmetry
algebras, it should be pointed out that the notion of spin
algebra, generalizing the standard notion of spin covering and
based on the division algebra structure of spinors alone, has been
introduced in \cite{fer}. On the other hand, a different
prescription for constructing generalized supersymmetries is also
possible and has been advocated in \cite{lt1}. It requires
matching the division algebra structures of both spinors and
Clifford algebras. According to the above table, e.g., in the
$5$-dimensional case a quaternionic structure can be imposed on
the supersymmetry since both spinors and Clifford algebras
arequaternionic. On the other hand in, let's say, the Minkowskian
$7$-dimensional case, at most a complex structure can be imposed,
because this is the minimal structure shared both by spinors and
Gamma matrices (see \cite{lt1} for details). We will come back
later on this issue. For the time being, let us just present
another table concerning the constraint generated by
division-algebra structures on generalized supersymmetries. For
the sake of clarity we will discuss fundamental spinors admitting
$32$ real components (as in the maximal supergravity or the
$11$-dimensional $M$-theory). Let us suppose that they admit a
real, complex, quaternionic\footnote{as this is the case,
e.g. for the $64$-component Euclidean
$D=11$ spinors. Quaternionic $32$-component spinors exist for instance in the $(3,7)$
signature.} or even an octonionic (as discussed later)
division algebra structure. Accordingly, the supersymmetry
generators $Q_a$ can be represented, respectively, as
$32$-dimensional real column vectors, $16$-dimensional complex,
$8$-dimensional quaternionic or $4$-dimensional octonionic
spinors. The generalized supersymmetry algebra
\begin{eqnarray}
\{ Q_a, {Q_b}^\ast \} &=& {\cal Z}_{ab},\label{gensusyalg}
\end{eqnarray}
where ${Q_a}^\ast$ denotes the principal conjugation in the given
division algebra, admits a hermitian r.h.s. (${\cal Z}_{ab} =
{\cal Z}_{ba}^\ast$), given by a hermitian matrix ${\cal Z}_{ab}$
of, respectively, $32\times 32$ real, $16\times 16$ complex,
$8\times 8$ quaternionic or $4\times 4$ octonionic-valued entries.
Due to the hermiticity condition, in the different cases, the
maximal number $\sharp$ of independent components for ${\cal
Z}_{ab}$ is given by { {{\begin{eqnarray}&
\begin{tabular}{|c|c|}\hline
  ${\bf \Psi}$&$\sharp$\\ \hline
${\bf R}~ (32) $& $528$\\ \hline ${\bf C}~ (16)$&$256$\\ \hline
 ${\bf H}~ ~(8)$&$120$\\ \hline
 ${\bf O} ~~(4)$ & $52$\\ \hline
\end{tabular}&\nonumber\end{eqnarray}}} }
It should be noticed that $528$ is the number of saturated
independent bosonic components in the $M$-algebra, deriving from
the real structure of $11$-dimensional Minkowskian spinors. As it
will be apparent in the following, an octonionic structure can be
imposed on Minkowskian $11$-dimensional spinors, leading to an
alternative, octonionic version of the $M$-algebra with only $52$
independent bosonic components.

\section{Clifford algebras revisited. Classification and explicit
constructions.}

For our purposes it is convenient to review the classification of
the irreducible representations of Clifford algebras from another
point of view, making explicit an algorithm allowing to single
out, in arbitrary signature space-times, a representative in each
irreducible class of representations of Clifford's gamma matrices.
As recalled in the previous section, the class of irreducible
representations is unique apart special signatures, where two
inequivalent irreducible representations are linked by sign
flipping ($\Gamma^\mu \leftrightarrow -\Gamma^\mu$). The explicit
construction presented here is the right tool allowing us to
introduce, in the next section, the quaternionic and octonionic
realizations for Clifford algebras and spinors.\par Our
construction goes as follows. At first we prove that starting from
a given $D$ spacetime-dimensional representation of Clifford's
Gamma matrices, we can recursively construct $D+2$ spacetime
dimensional Clifford  Gamma matrices with the help of two
recursive algorithms. Indeed, it is a simple exercise to verify
that if $\gamma_i$'s denotes the $d$-dimensional Gamma matrices of
a $D=p+q$ spacetime with $(p,q)$ signature (namely, providing a
representation for the $C(p,q)$ Clifford algebra) then
$2d$-dimensional $D+2$ Gamma matrices (denoted as $\Gamma_j$) of a
$D+2$ spacetime are produced according to either
\begin{eqnarray}
 \Gamma_j &\equiv& \left(
\begin{array}{cc}
  0& \gamma_i \\
  \gamma_i & 0
\end{array}\right), \quad \left( \begin{array}{cc}
  0 & {\bf 1}_d \\
  -{\bf 1}_d & 0
\end{array}\right),\quad \left( \begin{array}{cc}
  {\bf 1}_d & 0\\
  0 & -{\bf 1}_d
\end{array}\right)\nonumber\\
&&\nonumber\\ (p,q)&\mapsto&
 (p+1,q+1).\label{one}
\end{eqnarray}
or
\begin{eqnarray}
 \Gamma_j &\equiv& \left(
\begin{array}{cc}
  0& \gamma_i \\
  -\gamma_i & 0
\end{array}\right), \quad \left( \begin{array}{cc}
  0 & {\bf 1}_d \\
  {\bf 1}_d & 0
\end{array}\right),\quad \left( \begin{array}{cc}
  {\bf 1}_d & 0\\
  0 & -{\bf 1}_d
\end{array}\right)\nonumber\\
&&\nonumber\\ (p,q)&\mapsto&
 (q+2,p).\label{two}
\end{eqnarray}
It is immediate to notice, e.g., that the two-dimensional
real-valued Pauli matrices $\tau_A$, $\tau_1$, $\tau_2$ which
realize the Clifford algebra $C(2,1)$ are obtained by applying
either (\ref{one}) or (\ref{two}) to the number $1$, i.e. the
one-dimensional realization of $C(1,0)$. We have indeed
\begin{eqnarray}
&\tau_A= \left(\begin{array}{cc}0 &1\\ -1&0  \end{array}\right),
\quad \tau_1= \left(\begin{array}{cc}0 &1\\ 1&0
\end{array}\right), \quad
\tau_2= \left(\begin{array}{cc}1 &0\\ 0&-1  \end{array}\right).
\quad &\label{Pauli}
\end{eqnarray}
All Clifford algebras are obtained by recursively applying the
algorithms (\ref{one}) and (\ref{two}) to the Clifford algebra
$C(1,0)$ ($\equiv 1$) and the Clifford algebras of the series
$C(0, 3+4m)$ (with $m$ non-negative integer), which must be
previously known. This is in accordance with the scheme
illustrated in the table below.\par {~}\par {\em Table with the
maximal Clifford algebras (up to $d=256$).}
 {\tiny{
{\begin{eqnarray} &
\begin{tabular}{|cccccccccccccccccccc}
\hline
   1  &$\ast$& 2&$\ast$&  4&$\ast$& 8&$\ast$&16&$\ast$&32&$\ast$&64&
   $\ast$&128&$\ast$&256&$\ast$
      \\ \hline
  &&&&&&&&&&&&&\\
 $\underline{(1,0)}$ &$\Rightarrow$& $(2,1)$ &$\Rightarrow$&(3,2)
 &$\Rightarrow$&
  (4,3) &$\Rightarrow$&(5,4)& $\Rightarrow$ &(6,5)
&$\Rightarrow$&
  (7,6) &$\Rightarrow$&(8,7)& $\Rightarrow$ &(9,8)
  &$\Rightarrow$

  \\
  &&&&&&&&&&\\
  &&&&&&&&&&\\
   &&&&&&(1,4)&$\rightarrow $&(2,5)&$\rightarrow$&(3,6)
&$\rightarrow $&(4,7)&$\rightarrow$&(5,8)&$\rightarrow
$&(6,9)&$\rightarrow$
   \\
  &&&&&$\nearrow$&&&&&\\
  &&&&{\underline{(0,3)}}&& &&&&\\
  &&&&&$\searrow$&&&&&\\
  &&&&&&&&&&\\
  &&&&&&(5,0)&$\rightarrow
  $&(6,1)&$\rightarrow$&(7,2)&$\rightarrow$
&(8,3)&$\rightarrow $&(9,4)&$\rightarrow$&(10,5)&$\rightarrow$
  \\
   &&&&&&&&&&\\
  &&&&&&&&&&\\
   &&&&&&&&(1,8)&$\rightarrow $&(2,9)
&$\rightarrow $&(3,10)&$\rightarrow$&(4,11)&$\rightarrow
$&(5,12)&$\rightarrow$
   \\
  &&&&&&&$\nearrow$&&&\\
  &&&&&&{\underline{(0,7)}}&& &&\\
  &&&&&&&$\searrow$&&&\\
  &&&&&&&&&&\\
  &&&&&&&&(9,0)&$\rightarrow $&(10,1)
&$\rightarrow $&(11,2)&$\rightarrow$&(12,3)&$\rightarrow
$&(13,4)&$\rightarrow$
\\
&&&&&\\ &&&\\
  &&&& &&&&&&&&&&(1,12)&$\rightarrow $&(2,13)
&$\rightarrow $
   \\
  &&&&&&&&&&&&&$\nearrow$&&&\\
  &&&&&&&&&&&&{\underline{(0,11)}}&& &&\\
  &&&&&&&&&&&&&$\searrow$&&&\\
  &&&&&&&&&&&&&&\\
  &&&&&&&&&&&&&&(13,0)&$\rightarrow $&(14,1)
&$\rightarrow $\\ &&&&&\\ &&&\\
 &&&& &&&& &&&&&&&&(1,16)&$\rightarrow $
   \\
  &&&&&&&&&&&&&&&$\nearrow$&&&\\
  &&&&&&&&&&&&&&{\underline{(0,15)}}&& &&\\
  &&&&&&&&&&&&&&&$\searrow$&&&\\
  &&&&&&&&&&&&&&\\
  &&&&&&&&&&&&&&&&(17,0)&$\rightarrow $\\

\end{tabular}&\nonumber
\end{eqnarray}}}}
\begin{eqnarray}\label{bigtable}
&& \end{eqnarray} Concerning the previous table, some remarks are
in order. The columns are labeled by the matrix size ${d}$ of the
maximal Clifford algebras. Their signature is denoted by the
$(p,q)$ pairs. Furthermore, the underlined Clifford algebras in
the table can be named as ``primitive maximal Clifford algebras".
The remaining maximal Clifford algebras appearing in the table are
the ``maximal descendant Clifford algebras". They are obtained
from the primitive maximal Clifford algebras by iteratively
applying the two recursive algorithms (\ref{one}) and (\ref{two}).
Moreover, any non-maximal Clifford algebra is obtained from a
given maximal Clifford algebra by deleting a certain number of
Gamma matrices (as an example, Clifford algebras in
even-dimensional spacetimes are always non-maximal). \par It is
immediately clear from the above construction that the maximal
Clifford algebras are encountered if and only if the condition
\begin{eqnarray}
p -q &=& 1,5~ mod ~ 8
\end{eqnarray}
is matched. \par The notion of Clifford's algebra of Weyl type,
namely satisfying the (\ref{weyl}) condition, has already been
introduced. All maximal Clifford's algebras, both primitive and
descendant, are {\em not} of Weyl type. The non-maximal Clifford
algebras are of Weyl type if and only if they are produced from a
maximal Clifford algebra by deleting at least one temporal Gamma
matrix which, without loss of generality, can always be chosen the
one with diagonal entries. Explicitly, non-maximal Clifford
algebras are produced from the corresponding maximal Clifford
algebras according to the following table, which specifies the
number of time-like or space-like Gamma matrices that should be
deleted, as well as the Weyl ($W$) character or not ($NW$) of the
given non-maximal Clifford algebra. We get
\begin{eqnarray}&&
\begin{array}{|ccc||ccc|}\hline
  &$W$ & & &$NW$ &  \\ \hline
(0~mod~8)&\subset &(1~mod~8) &(2~mod~8) &\subset&(1~mod~8)\\
 (p,q) &\Leftarrow& (p+1,q)& (p,q)&\Leftarrow & (p,q+1)\\ \hline
 (4~mod~8)&\subset &(5~mod~8) &(3~mod~8) &\subset &(1~mod~8)\\
  (p,q) &\Leftarrow& (p+1,q)& (p,q)&\Leftarrow& (p,q+2)\\ \hline
  (6~mod~8)&\subset &(1~mod~8) & & &\\
  (p,q) &\Leftarrow& (p+3,q)&&&\\ \hline
  (7~mod~8)&\subset &(1~mod~8) & & &\\
   (p,q)&\Leftarrow& (p+2,q)&&&\\ \hline
\end{array}\label{realnmaxCl}
\end{eqnarray}
To complete our discussion what is left is specifying the
construction of the primitive maximal Clifford algebras for both
the $C(0, 3+8n)$ (which can be named as ``quaternionic series",
due to its connection with this division algebra, as we will see
in the next section), as well as the ``octonionic" series
$C(0,7+8n)$. The answer can be provided with the help of the three
Pauli matrices (\ref{Pauli}). We construct at first the $4\times
4$ matrices realizing the Clifford algebra $C(0,3)$ and the
$8\times 8$ matrices realizing the Clifford algebra $C(0,7)$. They
are given, respectively, by
\begin{eqnarray}
C(0,3) &\equiv& \begin{array}{c}
  \tau_A\otimes\tau_1, \\
  \tau_A\otimes\tau_2, \\
  {\bf 1}_2\otimes \tau_A.
\end{array}
\end{eqnarray}
and
\begin{eqnarray}
C(0,7) &\equiv& \begin{array}{c}
  \tau_A\otimes\tau_1\otimes{\bf 1}_2, \\
  \tau_A\otimes\tau_2\otimes{\bf 1}_2, \\
  {\bf 1}_2\otimes \tau_A\otimes \tau_1,\\
  {\bf 1}_2\otimes \tau_A\otimes \tau_2,\\
  \tau_1\otimes{\bf 1}_2\otimes\tau_A,\\
  \tau_2\otimes{\bf 1}_2\otimes\tau_A,\\
  \tau_A\otimes\tau_A\otimes\tau_A.
\end{array}\label{c07}
\end{eqnarray}
The three matrices of $C(0,3)$ will be denoted as ${\overline
\tau}_i$, $=1,2,3$. The seven matrices of $C(0,7)$ will be denoted
as ${\tilde \tau}_i$, $i=1,2,\ldots,7$. \par In order to construct
the remaining Clifford algebras of the two series we need at first
to apply the (\ref{one}) algorithm to $C(0,7)$ and construct the
$16\times 16$ matrices realizing $C(1,8)$ (the matrix with
positive signature is denoted as $\gamma_9$, ${\gamma_9}^2 ={\bf
1}$, while the eight matrices with negative signatures are denoted
as $\gamma_j$, $j=1,2\ldots , 8$, with ${\gamma_j}^2 =-{\bf 1}$).
We are now in the position to explicitly construct the whole
series of primitive maximal Clifford algebras $C(0,3+8n)$,
$C(0,7+8n)$ through the formulas
\begin{eqnarray}
C(0,3+8n)&\equiv& \begin{array}{lcr} {\overline\tau}_i\otimes
\gamma_9\otimes \ldots&\ldots&\ldots\otimes\gamma_9,\\ {\bf
1}_4\otimes\gamma_j\otimes{\bf 1}_{16}\otimes\ldots & \ldots &
\ldots\otimes{\bf 1}_{16},\\
 {\bf 1}_4\otimes\gamma_9\otimes\gamma_j\otimes {\bf
1}_{16}\otimes\ldots &\ldots&\ldots\otimes{\bf 1}_{16},  \\ {\bf
1}_4\otimes\gamma_9\otimes\gamma_9\otimes\gamma_j\otimes {\bf
1}_{16}\otimes \ldots&\ldots&\ldots\otimes{\bf 1}_{16},  \\ \ldots
&\ldots&\ldots, \\ {\bf
1}_4\otimes\gamma_9\otimes\ldots&\ldots&\otimes
\gamma_9\otimes\gamma_j,
\end{array}\label{quatern}
\end{eqnarray}
and similarly
\begin{eqnarray}
C(0,7+8n)&\equiv& \begin{array}{lcr} {\tilde\tau}_i\otimes
\gamma_9\otimes \ldots&\ldots&\ldots\otimes\gamma_9,\\ {\bf
1}_8\otimes\gamma_j\otimes{\bf 1}_{16}\otimes\ldots & \ldots &
\ldots\otimes{\bf 1}_{16},\\
 {\bf 1}_8\otimes\gamma_9\otimes\gamma_j\otimes {\bf
1}_{16}\otimes\ldots &\ldots&\ldots\otimes{\bf 1}_{16},  \\ {\bf
1}_8\otimes\gamma_9\otimes\gamma_9\otimes\gamma_j\otimes {\bf
1}_{16}\otimes \ldots&\ldots&\ldots\otimes{\bf 1}_{16},  \\ \ldots
&\ldots&\ldots, \\ {\bf
1}_8\otimes\gamma_9\otimes\ldots&\ldots&\otimes
\gamma_9\otimes\gamma_j,\label{octon}
\end{array}
\end{eqnarray}
Please notice that the tensor product of the $16$-dimensional
representation is taken $n$ times. The total size of the
(\ref{quatern}) matrix representations is then $4\times 16^n$,
while the total size of (\ref{octon}) is $8\times 16^n$.
\par
With the help of the formulas presented in this section we are
able to systematically construct a set of representatives of the
real irreducible representations of Clifford algebras in arbitrary
space-times and signatures.

\section{Quaternionic and octonionic realizations of Clifford
algebras.}

In this section we discuss the relations of Clifford algebras with
the division algebras of the quaternions (and of the octonions),
from a slightly different point of view w.r.t. the one expressed
in Section {\bf 2}.\par The relation can be understood as follows.
At first we notice that the three matrices appearing in $C(0,3)$
can also be expressed in terms of the imaginary quaternions
$\tau_i$ satisfying \begin{eqnarray} \tau_i\cdot\tau_j &=&
-\delta_{ij} +\epsilon_{ijk}\tau_k. \end{eqnarray} As a
consequence, the whole set of maximal primitive Clifford algebras
$C(0, 3+8n)$, as well as their maximal descendants, can be
represented with quaternionic-valued matrices. In its turn the
spinors have to be interpreted now as quaternionic-valued column
vectors.\par Similarly, there exists an alternative realization
for the Clifford algebra $C(0,7)$, obtained by identifying its
seven generators with the seven imaginary octonions satisfying the
algebraic relation
\begin{eqnarray}
\tau_i\cdot \tau_j &=& -\delta_{ij} + C_{ijk} \tau_{k},
\label{octonrel}
\end{eqnarray}
for $i,j,k = 1,\cdots,7$ and $C_{ijk}$ the totally antisymmetric
octonionic structure constants given by
\begin{eqnarray}
&C_{123}=C_{147}=C_{165}=C_{246}=C_{257}=C_{354}=C_{367}=1&
\end{eqnarray}
and vanishing otherwise. This octonionic realization of the
seven-dimensional Euclidean Clifford algebra will be denoted as
$C_{\bf O}(0,7)$. Due to the non-associative character of the
(\ref{octonrel}) octonionic product (the weaker condition of
alternativity is satisfied, see \cite{gk}), the octonionic
realization cannot be represented as an ordinary matrix product
and is therefore a distinct and inequivalent realization of this
Euclidean Clifford algebra with respect to the one previously
considered (\ref{c07}). Please notice that, by iteratively
applying the two lifting algorithms to $C_{\bf O}(0,7)$, we obtain
matrix realizations (with octonionic-valued entries) for the
maximal Clifford algebras of the series $C(n, 7+n)$ and $C(8+n,
n-1)$, for positive integral values of $n$ ($n=1,2,\ldots$). The
dimensionality of the corresponding octonionic-valued matrices is
$2^n\times 2^n$. For completeness we should point out that the
construction (\ref{octon}) leading to the primitive maximal
Clifford algebras $C(0, 7+8n)$, can be carried on with the help of
an octonionic-valued realization of the $\gamma_9$ matrix. As a
consequence, realizations of $C(0,7+8n)$ and their descendants can
be produced acting on column spinors, whose entries are tensor
products of octonions. In any case, in the following, we will
focus on the single octonionic realizations $C_{\bf O} (n, 7+n)$
and $C_{\bf O}(9+n, n)$ (here $n=0,1,2,\ldots $) which are of
relevance in the context of the $M$-theory.
\par
One should be aware of the properties of the non-associative
realizations of Clifford algebras. In the octonionic case the
commutators $\Sigma_{\mu\nu} =\relax [\Gamma_\mu, \Gamma_\nu]$ are
no longer the generators of the Lorentz group. They correspond
instead to the generators of the coset $SO(p,q)/G_2$, being $G_2$
the $14$-dimensional exceptional Lie algebra of automorphisms of
the octonions. As an example, in the Euclidean $7$-dimensional
case, these commutators give rise to $7=21-14$ generators,
isomorphic to the imaginary octonions. Indeed
\begin{eqnarray}
\relax [\tau_i,\tau_j]& = &2C_{ijk}\tau_k .\label{octcomm}
\end{eqnarray}
The alternativity property satisfied by the octonions implies that
the seven-dimensional commutator algebra among imaginary octonions
is not a Lie algebra, the Jacobi identity being replaced by a
weaker condition that endorses (\ref{octcomm}) with the structure
of a Malcev algebra (see \cite{gk}).\par Such an algebra admits a
nice geometrical interpretation \cite{{hl},{cp}}. Indeed, the
normed $1$ unitary octonions $X=x_0 + x_i\tau_i$ (with $x_0$ and
$x_i$, for $i=1,\ldots,7$, real and the summation over repeated
indices understood), i.e. restricted by the condition
\begin{eqnarray}
X^\dagger\cdot X&=&1,\label{uninorm}
\end{eqnarray}
describe the seven-sphere $S^7$. The latter is a parallelizable
manifold with a quasi (due to the lack of associativity) group
structure. Here $X^\dagger$ denotes the principal conjugation for
the octonions, namely \begin{eqnarray} X^\dagger &= & x_0
-x_i\tau_i. \label{prconj}\end{eqnarray}
On the seven sphere, infinitesimal
homogeneous transformations which play the role of the Lorentz
algebra can be introduced through
\begin{eqnarray}
\delta X &=& a\cdot X,\end{eqnarray} with $a$ an infinitesimal
constant octonion. The requirement of preserving the unitary norm
(\ref{uninorm}) implies the vanishing of the $a_0$ component, so
that $a \equiv a_i\tau_i$. Therefore, the above commutator algebra
(\ref{octcomm}), generated by the seven $\tau_i$, can be
interpreted as the algebra of ``quasi" Lorentz transformations
acting on the seven sphere $S^7$. At least in this specific
example we discovered a nice geometrical setting underlining the
use of the octonionic realization of the $C_{\bf O}(0,7)$ Clifford
algebra. While the associative (\ref{c07}) representation of the
seven dimensional Clifford algebra is required for describing the
Euclidean $7$-dimensional flat space, the non-associative
realization describes the geometry of $S^7$.

\section{On real, quaternionic and octonionic spinors.}

In this section we introduce (following \cite{kt}, where real and
complex spinors were treated), the necessary ingredients and
conventions to introduce the spinorial dynamics. Quaternionic and
octonionic spinors are considered as well.\par In \cite{kt} three
matrices (only two of them independent) $A$, $B$, $C$, associated
to the three conjugations (hermitian, complex and transposition)
acting on Gamma matrices, were introduced. In the case of the
restriction to real-valued Gamma matrices, only one matrix
(conventionally denoted as $A$, see \cite{dt}) needs to be
introduced. $A$ plays the role of $\Gamma^0$ in the Minkowskian
case and serves to introduce  barred spinors. In a $(t,s)$
spacetime $A$ is, up to a sign, the product of the time-like Gamma
matrices and satisfies the relations
\begin{eqnarray}
A\Gamma^\mu A^T &=& \xi {\Gamma^\mu}^T,\nonumber\\ A^T &=& \alpha
A,
\end{eqnarray}
with
\begin{eqnarray}
\xi &=& (-1)^{t-1}, \nonumber \\ \alpha  &=& (-1)^{t(t-1)/2},
\label{alphaxi}\end{eqnarray}
as it can be easily checked.\par In both the quaternionic and
octonionic case, two real-valued matrices, conventionally denoted
as $A$ and $C$, can be introduced. As before, $A$ plays the role
of $\Gamma^0$ and is used to define barred spinors (${\overline
\psi}= \psi^\dagger A$). $A$ and $C$ satisfy the set of relations
\begin{eqnarray}
A\Gamma_{\mu} A^{\dagger}  &=& \xi \Gamma_{\mu}^{\dagger},
\nonumber \\ C\Gamma_{\mu} C^{\dagger}  &=& \delta
\Gamma_{\mu}^{T}, \nonumber \\ C^{T} &=& \rho C, \nonumber\\
A^{\dagger} &=& \alpha A, \nonumber \\ A^{T} &=& \sigma C A
C^{\dagger},\label{signs}
\end{eqnarray}
where ``$\dagger$" denotes the combination of matrix transposition
and principal conjugation in the division algebra (see
(\ref{prconj})). The signs $\alpha$, $\xi$, $\delta$, $\rho$,
$\sigma$ will be specified below.\par The matrix $A$ is always
given by the product of the temporal $\Gamma$'s (regardless of the
order), while up to two inequivalent $C$ matrices can be found,
given by the product (again, regardless of the order) of
respectively all symmetric ($C_S$) or all antisymmetric ($C_A$)
Gamma matrices (in special cases $C_S$, $C_A$ collapse to the
single matrix $C$).\par For maximal Clifford algebras (in the
sense specified in Section {\bf 3}) of a $(t,s)$ space-time, the
set of signs is given by
\begin{eqnarray}
\alpha  &=& (-1)^{t(t-1)/2}, \nonumber \\ \xi &=& (-1)^{t-1},
\nonumber \\ \delta &=& (-1)^{t} \nonumber \\ \rho &=&
(-1)^{t(t+1)/2}, \nonumber \\ \sigma &=& \sin( \frac{|t-s|\pi}{2})
(-1)^{\frac{t(t+1)}{2}+1},
\end{eqnarray}
as it can be checked with straightforward computations. Please
notice that the matrix $C$ is unique in this case.
\par The maximal quaternionic Clifford algebras are those
satisfying the
\begin{eqnarray}
t-s &=& 5~ mod ~8
\end{eqnarray}
condition, while the maximal octonionic Clifford algebras are the
subclass of
\begin{eqnarray}
t-s&=& 1~mod~8
\end{eqnarray}
maximal Clifford algebras, obtained after erasing the series
corresponding to the first row in table (\ref{bigtable}) (i.e.
$t=s+1$).\par Just like the real case, non-maximal Clifford
algebras are obtained after erasing a certain number of Gamma
matrices. The quaternionic equivalent of table (\ref{realnmaxCl})
is given, in the quaternionic case, by
\begin{eqnarray}&&
\begin{array}{|ccc||ccc|}\hline
  &$W$ & & &$NW$ &  \\ \hline
(4~mod~8)&\subset &(5~mod~8) &(6~mod~8) &\subset&(5~mod~8)\\
 (p,q) &\Leftarrow& (p+1,q)& (p,q)&\Leftarrow & (p,q+1)\\ \hline
 (3~mod~8)&\subset &(5~mod~8) &(7~mod~8) &\subset &(5~mod~8)\\
  (p,q) &\Leftarrow& (p+2,q)& (p,q)&\Leftarrow& (p,q+2)\\ \hline
  (2~mod~8)&\subset &(5~mod~8) &(0~mod~8) &\subset &(5~mod~8)\\
  (p,q) &\Leftarrow& (p+3,q)&(p,q)&\Leftarrow&(p,q+3)\\ \hline
  (1~mod~8)&\subset &(5~mod~8) & & &\\
   (p,q)&\Leftarrow& (p+4,q)&&&\\ \hline
\end{array}
\end{eqnarray}
while, in the octonionic case, we have the table
\begin{eqnarray}&&
\begin{array}{|ccc||ccc|}\hline
  &$W$ & & &$NW$ &  \\ \hline
(0~mod~8)&\subset &(1~mod~8) &(2~mod~8) &\subset&(1~mod~8)\\
 (p,q) &\Leftarrow& (p+1,q)& (p,q)&\Leftarrow & (p,q+1)\\ \hline
 (7~mod~8)&\subset &(1~mod~8) &(3~mod~8) &\subset &(1~mod~8)\\
  (p,q) &\Leftarrow& (p+2,q)& (p,q)&\Leftarrow& (p,q+2)\\ \hline
  (6~mod~8)&\subset &(1~mod~8) &(4~mod~8) &\subset &(1~mod~8)\\
  (p,q) &\Leftarrow& (p+3,q)&(p,q)&\Leftarrow&(p,q+3)\\ \hline
  (5~mod~8)&\subset &(1~mod~8) & & &\\
   (p,q)&\Leftarrow& (p+4,q)&&&\\ \hline
\end{array}
\end{eqnarray}
We should mention that, to be consistent, in, let's say, the
octonionic realization of a non-maximal Clifford algebra, all the
seven matrices proportional to the imaginary octonions must be
present. Stated otherwise, the deleted matrices from the
corresponding maximal Clifford algebra are all real-valued.\par
For completeness, let us right down the values of the signs
entering (\ref{signs}) for the quaternionic and octonionic
non-maximal Clifford algebra cases obtained by deleting a single
Gamma matrix. In all four cases below two inequivalent $C$
matrices are present and the suffix ($S$ or$A$) specifies whether
$C_S$ or $C_A$ is involved, while the signs $\alpha$, $\xi$ are
given by (\ref{alphaxi}). Furthermore, in all four cases below we
get
\begin{eqnarray}
 \delta_{S} &=& (-1)^{t}, \nonumber
\\ \delta_{A} &=& (-1)^{t+1}.
\end{eqnarray}
The remaining signs are given by \par {\em i}) {\em in the
quaternionic $4~mod~8$ (W) case},

\begin{eqnarray}
\rho_{S} &=&  (-1)^{t(t+1)/2}, \nonumber
\\ \rho_{A} &=& - (-1)^{t(t-1)/2}, \nonumber
\\ \sigma_{S} &=&
\sin((t-s)\frac{3\pi}{8})(-1)^{\frac{t(t+1)}{2}}, \nonumber \\
\sigma_{A} &=& \sin((t-s)\frac{3\pi}{8})(-1)^{\frac{t(t-1)}{2}},
\nonumber
\end{eqnarray}

{\em ii}) {\em in the quaternionic $6~mod~8$ (NW) case},

\begin{eqnarray}
\rho_{S} &=&  (-1)^{t(t+1)/2}, \nonumber
\\ \rho_{A} &=& (-1)^{t(t-1)/2},
\nonumber
\\ \sigma_{S} &=& \sin (|t-s|3\pi/4) (-1)^{t(t+1)/2 +1}, \nonumber
\\ \sigma_{A} &=& \sin (|t-s|3\pi/4) (-1)^{t(t-1)/2 +1}, \nonumber
\end{eqnarray}

{\em iii}) {\em in the octonionic $0~mod~8$ (W) case},

\begin{eqnarray}
 \rho_{S} &=&  (-1)^{t(t+1)/2}, \nonumber
\\ \rho_{A} &=& -(-1)^{t(t-1)/2}, \nonumber
\\
\sigma_{S} &=& \sin((t-s)\frac{3\pi}{16}) (-1)^{t(t+1)/2},
\nonumber
\\ \sigma_{A} &=& \sin((t-s)\frac{3\pi}{16}) (-1)^{t(t-1)/2},
\nonumber
\end{eqnarray}

{\em iv}) {\em and finally in the octonionic $2~mod~8$ (NW) case},

\begin{eqnarray}
 \rho_{S} &=& (-1)^{t(t+1)/2}, \nonumber
\\ \rho_{A} &=& (-1)^{t(t-1)/2}, \nonumber
\\
\sigma_{S} &=& \sin(|t-s|\frac{\pi}{4}) (-1)^{\frac{t(t+1)}{2}
+1}, \nonumber \\ \sigma_{A} &=& \sin(|t-s|\frac{\pi}{4})
(-1)^{\frac{t(t-1)}{2} +1}. \nonumber
\end{eqnarray}

We remind that in the Weyl ($W$) case, the projectors $P_\pm$ can
be introduced through
\begin{eqnarray}
P_\pm &=&\frac{1}{2}({\bf 1}_{2d} \pm {\overline
\Gamma}),\nonumber\\
{\overline\Gamma} &=&\left( \begin{array}{cc}
  {\bf 1}_d & 0 \\
  0 & -{\bf 1}_d
\end{array}\right)\label{project}
\end{eqnarray}
and chiral (antichiral) spinors can be defined through
\begin{eqnarray}
\Psi_\pm &=& P_\pm \Psi.
\end{eqnarray}
It is worth ending this section writing down, symbolically, the
most general spinorial terms in a free lagrangian which can
possibly (depending on the signature and dimensionality of the
space-time) be encountered in our theories. It is sufficient to
list such terms in the octonionic case. One trivially realizes how
to employ the same symbols in the quaternionic and real cases as
well.\par Different massive terms can be found in the Weyl ($W$)
case, i.e.\footnote{here ``$tr$" denotes the projection onto the
octonionic identity, $ tr(x_0+x_i\tau_i)=x_0$. It coincides with
the standard trace when we are restricting to the quaternionic
subcase.}
\begin{eqnarray}
M_{//} &=& tr(\Psi^{\dagger}_{+} A \Psi_{+}), \nonumber \\
M_{\bot} &=& tr(\Psi^{\dagger}_{+} A \Psi_{-} + \Psi^{\dagger}_{-}
A \Psi_{+}),\nonumber \\ M_{//T,S} &=& tr(\Psi^{\dagger}_{+} A
\Gamma_{T,S} \Psi_{+}), \nonumber \\ M_{\bot T,S} &=&
tr(\Psi^{\dagger}_{+} A \Gamma_{T,S}\Psi_{-} + \Psi^{\dagger}_{-}
A \Gamma_{T,S}\Psi_{+}),\nonumber \\ M_{//J} &=&
tr(\Psi^{\dagger}_{+} A J \Psi_{+}), \nonumber \\ M_{\bot J} &=&
tr(\Psi^{\dagger}_{+} A J \Psi_{-} + \Psi^{\dagger}_{-} A J
\Psi_{+}),\nonumber\\
 M_{//F} &=& tr(\Psi^{\dagger}_{+} A F \Psi_{+}),\nonumber \\
 M_{\bot F} &=&  tr(\Psi^{\dagger}_{+} A F \Psi_{-}+ \Psi^{\dagger}_{-} A F \Psi_{+}),
\label{bilinmass}
\end{eqnarray}
where $\Gamma_T$, $\Gamma_S$ denote, in a non-maximal Clifford
algebra case, the presence of an external (deleted from the set of
maximal Gamma's) Gamma matrix of time, or respectively, space-like
type. Similarly, $J$ denotes the product of two such matrices,
either time-like or space-like, while $F$ denotes the product of three
external matrices. No other massive symbols need to
be introduced, as it will appear from the tables given in the
following. In a $NW$-case, similar symbols can be introduced.
However, since in this case no chiral (antichiral) spinors are
defined, full spinors are present in the r.h.s. and the ``$//$"
and ``$\bot$" suffices must be dropped.\par In full analogy, the
set of symbols in a Weyl ($W$) kinetic case are given
by\footnote{as before, analogous symbols are employed in the
$NW$-case, by dropping the suffices ``$//$" and ``$\bot$".}
\begin{eqnarray} K_{//} &=& \frac{1}{2}tr[(\Psi^{\dagger}_{+} A
\Gamma^{\mu})\partial_{\mu} \Psi_{+}]+
\frac{1}{2}tr[\Psi^{\dagger}_{+} (A \Gamma^{\mu}\partial_{\mu}
\Psi_{+})], \nonumber \\ K_{\bot} &=&\frac{1}{2}
tr[(\Psi^{\dagger}_{+} A \Gamma^{\mu})\partial_{\mu} \Psi_{-}] +
\frac{1}{2}tr[\Psi^{\dagger}_{+} (A \Gamma^{\mu}\partial_{\mu}
\Psi_{-})] + \nonumber\\ &&\frac{1}{2} tr[(\Psi^{\dagger}_{-} A
\Gamma^{\mu})\partial_{\mu} \Psi_{+}] +
\frac{1}{2}tr[\Psi^{\dagger}_{-} (A \Gamma^{\mu}\partial_{\mu}
\Psi_{+})], \nonumber \\ K_{//T,S} &=&
\frac{1}{2}tr[(\Psi^{\dagger}_{+} A \Gamma^{\mu}
\Gamma_{T,S})\partial_{\mu} \Psi_{+}]+
\frac{1}{2}tr[\Psi^{\dagger}_{+} (A \Gamma^{\mu}
\Gamma_{T,S}\partial_{\mu} \Psi_{+})], \nonumber \\ K_{\bot T,S}
&=&\frac{1}{2} tr[(\Psi^{\dagger}_{+} A
\Gamma^{\mu}\Gamma_{T,S})\partial_{\mu} \Psi_{-}] +
\frac{1}{2}tr[\Psi^{\dagger}_{+} (A
\Gamma^{\mu}\Gamma_{T,S}\partial_{\mu} \Psi_{-})] +\nonumber\\ &&
\frac{1}{2} tr[(\Psi^{\dagger}_{-} A
\Gamma^{\mu}\Gamma_{T,S})\partial_{\mu} \Psi_{+}] +
\frac{1}{2}tr[\Psi^{\dagger}_{-} (A
\Gamma^{\mu}\Gamma_{T,S}\partial_{\mu} \Psi_{+})] \nonumber \\
K_{//J} &=& \frac{1}{2}tr[(\Psi^{\dagger}_{+} A \Gamma^{\mu}
J)\partial_{\mu} \Psi_{+}]+ \frac{1}{2}tr[\Psi^{\dagger}_{+} (A
\Gamma^{\mu} J\partial_{\mu} \Psi_{+})], \nonumber \\ K_{\bot J}
&=&\frac{1}{2} tr[(\Psi^{\dagger}_{+} A \Gamma^{\mu}
J)\partial_{\mu} \Psi_{-}] + \frac{1}{2}tr[\Psi^{\dagger}_{+} (A
\Gamma^{\mu} J\partial_{\mu} \Psi_{-})] +\nonumber\\&& \frac{1}{2}
tr[(\Psi^{\dagger}_{-} A \Gamma^{\mu} J)\partial_{\mu} \Psi_{+}]
+\frac{1}{2}tr[\Psi^{\dagger}_{-} (A \Gamma^{\mu} J\partial_{\mu}
\Psi_{+})], \nonumber \\
 K_{//F} &=& \frac{1}{2}tr[(\Psi^{\dagger}_{+} A \Gamma^{\mu} F)\partial_{\mu} \Psi_{+}]+ \frac{1}{2}tr[\Psi^{\dagger}_{+} (A \Gamma^{\mu} F\partial_{\mu} \Psi_{+})], \nonumber \\
  K_{\bot F} &=&\frac{1}{2} tr[(\Psi^{\dagger}_{+} A \Gamma^{\mu}
 F)\partial_{\mu} \Psi_{-}] + \frac{1}{2}tr[\Psi^{\dagger}_{+} (A
 \Gamma^{\mu} F\partial_{\mu} \Psi_{-})] + \nonumber\\&&\frac{1}{2}
 tr[(\Psi^{\dagger}_{-} A \Gamma^{\mu} F)\partial_{\mu} \Psi_{+}]
 +
 \frac{1}{2}tr[\Psi^{\dagger}_{-} (A \Gamma^{\mu} F\partial_{\mu} \Psi_{+})] .
\end{eqnarray}
Please notice that, due to the non-associativity of the octonions,
in the kinetic case we have to correctly specify the order in
which the operations are taken. There is no such problem in the
massive case since the matrices $\Gamma_T$, $\Gamma_S$, $J$ and $F$ can
always be chosen, without loss of generality, real. Therefore in
(\ref{bilinmass}) at most bilinear octonionic terms are present and the
non-associativity of the octonions plays no role.

\section{The real case revisited.}

In reference {\cite{dt} the Majorana condition for complex
spinors was analyzed and the list of different signature
spacetimes allowing for kinetic, pseudokinetic, massive and/or
pseudomassive terms in the free-Majorana spinors lagrangians were
presented. A slight generalization of these results can be
produced in this section, based of the classification of real
spinors that we have previously discussed (we notice,{\em en
passant}, that the spinors we are dealing with here are, by
construction, real, so that no Majorana condition, referring to a
previous complex structure, needs to be imposed).\par It is just a
matter of lengthy, but straightforward computations, to produce a
set of tables of the allowed, non-vanishing, free kinetic and
massive terms in each given signature space-times. In the
following tables, the columns are labeled by $t~mod~4$, while the
rows by $t-s~mod~8$. The entries represent, simbolically, the
allowed kinetic and/or massive terms (the precise meaning of the
symbols is discussed at the end of the previous section). An empty
space means that, neither a kinetic, nor a massive term is allowed
for the corresponding space-time.\par The first table is produced
for the real $NW$ case. We get \begin{eqnarray}&&
{\begin{tabular}{||c||c|c|c|c|c|} \hline \hline
 & 0 & 1 & 2 & 3 \\ \hline
    \hline
  1  &  &  K & K,M &  M  \\ \hline
   2 & $ M_{S}$ & K &  K, $K_{S}$, M & $K_{S}$, M, $M_{S}$ \\ \hline
  3 & $M_{S1}$, $M_{S2}$, $M_{J}$, $K_{J}$  & K,$M_{J}$ & K,
  $K_{S1}$,$K_{S2}$,M &  $K_{S1}$,$K_{S2}$,$K_{J}$,M, $M_{S1}$,$M_{S2}$  \\  \hline
  5 &   & K & K,M &  M  \\  \hline
 \end{tabular}}\end{eqnarray}

The second table is for the real $W$ (Weyl) case. We have in this
case \begin{eqnarray}&&
 {\begin{tabular}{||c||c|c|c|c|c|} \hline \hline
    & 0 & 1 & 2 & 3 \\ \hline
    \hline
 0 &  & $K_{//}$  & $M_{//}$, $K_{\bot}$ &  $M_{\bot}$   \\  \hline
 4 &   & $K_{//}$ & $M_{//}$, $K_{\bot}$  & $M_{\bot}$  \\ \hline
 6 & $K_{// T1}$, $K_{// T2}$, $M_{// J}$,  & $K_{//}$, $M_{// T1}$,$M_{// T2}$, &
  $M_{//}$, $K_{\bot}$, $M_{\bot T1}$,  & $K_{// J}$, $M_{\bot}$   \\
   & $K_{\bot J}$ & $K_{\bot T1}$, $K_{\bot T2}$, $M_{\bot J}$ & $M_{\bot T2}$ & \\ \hline
7  & $K_{//T}$ & $K_{//}$, $M_{//T}$,$K_{\bot T}$   & $M_{//}$,
  $K_{\bot}$, $M_{\bot T}$ & $M_{\bot}$    \\  \hline
 \end{tabular}}\end{eqnarray}

\section{Quaternionic spinors and their free dynamics. A
classification.}

In this section we present the tables of allowed free kinetic and
massive terms for quaternionic spinors. As in the previous
section, the columns are labeled by $t~mod~4$ and the rows by
$t-s~mod~8$, while the symbols used in the entries are explained
at the end of section {\bf 5}.\par In the $NW$ case we have
\begin{eqnarray}&&
{\begin{tabular}{||c||c|c|c|c|c|} \hline \hline
   & 0 & 1 & 2 & 3 \\ \hline
    \hline
   0&$K_{J_j}$, $K_F$, $M_{S_j}$, $M_{J_j}$  &$K$, $K_F$, $M_{J_j}$&
$K$, $K_{S_j}$, $M$, $M_F$ &$K_{S_j}$, $K_{J_j}$, $M$, $M_{S_j}$, $M_F$ \\ \hline
  5  &  &  $K$ & $K$, $M$ &  $M$  \\ \hline
  6 & $M_{S}$  & $K$  & $K$, $K_{S}$, $M$ &  $K_{S}$, $M$, $M_{S}$  \\  \hline
  7 &$ K_{J}$, $ M_{S_i}$,$ M_{J}$ & $K$, $ M_{J}$ &
  $K$, $K_{S_i}$, $M$ & $K_{S_i}$ ,$ K_{J}$, $M$, $M_{S_i}$ \\ \hline

 \end{tabular}}\nonumber\end{eqnarray}
\begin{eqnarray}
&&
\end{eqnarray}

In the $W$ (Weyl) case we have
\begin{eqnarray}&&
{\begin{tabular}{||c||c|c|c|c|c|} \hline \hline
   & 0 & 1 & 2 & 3 \\ \hline
    \hline
  1&$
\begin{array}{l} K_{// T_j}, K_{\bot J_j},\\ M_{//J_j}
\end{array}$& $\begin{array}{l} K_{//},
K_{\bot T_j},\\ M_{// F}, M_{//J_j}, M_{\bot J_j}\end{array}$&
$\begin{array}{l}K_{// F}, K_\bot,\\ M_{//}, M_{\bot F}, M_{\bot T_j}\end{array}$&
$\begin{array}{l} K_{\bot F}, K_{// J_j},\\ M_\bot \end{array}$
\\ \hline
  2 & $\begin{array}{l} K_{// T_i}, K_{\bot J},\\ M_{// J}\end{array} $ &
$\begin{array}{l} K_{//}, K_{\bot T_i},\\
  M_{// T_i}, M_{\bot J}\end{array}$ &
$\begin{array}{l} K_{\bot},\\ M_{//},M_{\bot T_i}\end{array}$  & $\begin{array}{l} K_{// J},\\ M_{\bot}\end{array}$   \\ \hline
  3  & $\begin{array}{l}K_{//T}\\ \end{array}$ & $\begin{array}{l}K_{//}, K_{\bot T},\\
 M_{//T}\end{array}$   & $\begin{array}{l} K_\bot , \\M_{//}, M_{\bot T}\end{array}$
  & $\begin{array}{l} \\ M_{\bot}\end{array}$    \\  \hline
  4 &  & $\begin{array}{l} K_{//}\\ \end{array}$  & $\begin{array}{l} K_\bot , \\
M_{//}\end{array}$  &  $\begin{array}{l}\\M_{\bot}\end{array}$   \\  \hline
 \end{tabular}}\nonumber\end{eqnarray}
\begin{eqnarray}
&&
\end{eqnarray}
Please notice that in the two tables above the suffix ``$j$" denotes the existence of three inequivalent choices for the corresponding matrices (e.g., the three distinct space-like matrices $S_j$),
while the suffix ``$i$" denotes the existence of two inequivalent choices. As previously discussed, this is
in accordance with the signature of the given space-time. Therefore, the let's say, $t-s=0~mod~8$, $t=2~mod~4$ spacetime
admits, besides the $K$ kinetic term, three extra kinetic terms $K_{S_j}$ associated to the three external space-like Gamma matrices $S_j$,
$j=1,2,3$, existing in this case.

\section{Octonionic spinors and their free dynamics. A
classification.}

Here we present the tables of allowed free kinetic and massive
terms for octonionic spinors. As in the two previous sections, the
columns are labeled by $t~mod~4$ and the rows by $t-s~mod~8$,
while the symbols used in the entries are explained at the end of
section {\bf 5}.\par In the $NW$ case we have
\begin{eqnarray}&&
{\begin{tabular}{||c||c|c|c|c|c|} \hline \hline
   & 0 & 1 & 2 & 3 \\ \hline
    \hline
  1  &  &  $K$ & $K$, $M$ &  $M$  \\ \hline
  2 & $M_{S}$  & $K$  & $K$, $K_{S}$, $M$ &  $K_{S}$, $M$, $M_{S}$  \\  \hline
  3 & $ K_{J}$,$ M_{S_i}$,$ M_{J}$ & $K$, $ M_{J}$ &
  $K$, $K_{S_i}$, $M$ & $K_{S_i}$, $ K_{J}$, $M$, $M_{S_i}$ \\ \hline
4 & $K_{J_j}$, $K_F$, $M_{S_j}$, $M_{J_j}$& $K$, $K_F$, $M_{J_j}$, $M_F$&
$K$, $K_{S_j}$, $M$, $M_F$& $K_{S_j}$, $K_{J_j}$, $M$, $M_{S_j}$\\ \hline
 \end{tabular}}\nonumber\end{eqnarray}
\begin{eqnarray}
&&
\end{eqnarray}
In the $W$ (Weyl) case we have
\begin{eqnarray}&&
{\begin{tabular}{||c||c|c|c|c|c|} \hline \hline
   & 0 & 1 & 2 & 3 \\ \hline
    \hline
  0 &  & $\begin{array}{l} K_{//}\\ \end{array}$  & $\begin{array}{l}
K_\bot ,\\ M_{//}\end{array}$ &$\begin{array}{l} \\  M_{\bot}\end{array}$   \\  \hline
5 & $\begin{array}{l} K_{//T_j}, K_{\bot J_{j}},\\ M_{// J_j}, M_{\bot F} \end{array}$
&
$\begin{array}{l} K_{//}, K_{\bot T_j}, \\ M_{// T_j}, M_{\bot J_j}\end{array}$&
$\begin{array}{l} K_{\bot} , K_{//F}, \\ M_{//}, M_{\bot T_j}\end{array}$&
$\begin{array}{l} K_{// J_j}, K_{\bot F}, \\ M_{\bot}, M_{// F} \end{array}$\\ \hline
  6 & $\begin{array}{l}K_{// T_i}, K_{\bot J},\\  M_{// J},\end{array}$  & $\begin{array}{l}
  K_{//}, K_{\bot T_i}, \\
  M_{// T_i}, M_{\bot J}\end{array}$ & $\begin{array}{l} K_\bot , \\M_{//},
  M_{\bot T_i}\end{array}$  & $\begin{array}{l} K_{// J},\\ M_{\bot}\end{array}$   \\
  \hline
  7  & $\begin{array}{l}K_{//T}\\
\end{array}$ & $\begin{array}{l}K_{//},K_{\bot T},\\ M_{//T}\end{array}$  &
$\begin{array}{l} K_\bot ,\\M_{//}, M_{\bot T}\end{array}$
 & $\begin{array}{l} \\M_{\bot}\end{array}$    \\  \hline
 \end{tabular}}\nonumber\end{eqnarray}
\begin{eqnarray}
&& \end{eqnarray} As in the previous section, the suffices ``$i$"
and ``$j$" takes two and respectively three distinct values. With
these last tables we completed the classification of the allowed
free lagrangians for spinors in different space-times.

\section{Identities for higher rank antisymmetric octonionic tensors.}

As we have seen in the previous sections, octonionic spinors are
associated with octonionic Clifford algebras. In their turn, these
ones are given by the maximal octonionic Clifford algebras,
specified by the two sets of octonionic realizations for the
signatures
\begin{eqnarray}
C_{\bf O}(m, 7+8n+m) &,& C_{\bf O}(9+8n+m,m),\label{octonmax}
\end{eqnarray}
with $n,m\geq 0$, together with the class of octonionic
non-maximal Clifford algebras obtained from (\ref{octonmax}) by
deleting a certain number of {\em real-valued} Gamma matrices. The
reality restriction on these extra Gamma matrices (which cannot
therefore contain imaginary octonions) puts a constraint on the
space-time signatures admitting an octonionic description. For
later convenience, it is useful to present the list of the whole
class of octonionic space-times recovered from the maximal
Clifford algebras of space-time dimension $D=t+s$ up to $D=13$.
The following table can be produced, with the columns labeled by
$D$, the dimensionality of the spacetime. The maximal Clifford
algebras are underlined. In each entry the octonionic
dimensionality ${\bf d_{\Psi}}$ of the fundamental spinors is also
reported. The signatures admitting, for each given spacetime
dimension $D$, spinors of minimal octonionic dimensionality are
denoted with a ``$\ast$". Finally, the chain of reductions from a
given maximal Clifford algebra is sketchily reported (please
notice that the chain of reductions is not necessarily unique, for
instance the $(10,1)$ signature can be produced by erasing a
single Gamma matrix either from $(11,1)$ or from $(10,2)$). We get
\begin{eqnarray}&&
\begin{array}{|c|c|c|c|c|c|c|}\hline
7&8&9&10&11&12&13\\ \hline
 {\underline{(0,7)}}^\ast, {\bf 1} && & & & & \\ \hline
(7,0)^{\ast}, {\bf 1} &(8,0)^\ast , {\bf 1} &{\underline{(9,0)}}^{\ast},
{\bf 2}&&&&\\ \hline & (0,8)^\ast ,{\bf 1}&
{\underline{(1,8)}}^\ast , {\bf 2} &&&&\\ \hline &
\begin{array}{c}\\(7,1), {\bf 2}\end{array}& \begin{array}{c}\\(8,1)^\ast ,{\bf 2}\end{array}
& \begin{array}{c} (10,0), {\bf 4}\\(9,1)^\ast , {\bf
2}\end{array}&\underline{(10,1)}^\ast ,{\bf 4}&&\\ \hline

 &&
\begin{array}{c}(0,9)^\ast , {\bf 2}\\(2,7),{\bf 4}\end{array}& \begin{array}{c}(1,9)^\ast ,
{\bf 2}\\(2,8),{\bf 4}
\end{array}
&\underline{(2,9)}^\ast ,{\bf 4}&&\\
 \hline

  &&
\begin{array}{c}\\(7,2), {\bf 4}\end{array}& \begin{array}{c}\\(8,2),{\bf 4}\end{array}
&
\begin{array}{c}(11,0),{\bf 8} \\(9,2)^\ast,{\bf 4}\end{array}
& \begin{array}{c}(11,1), {\bf 8}\\(10,2)^\ast , {\bf
4}\end{array}&\underline{(11,2)}^\ast ,{\bf 8}\\
 \hline

  &&&
\begin{array}{c} (0,10), {\bf 4}\\(3,7),{\bf 8}\end{array}& \begin{array}{c}(1,10)^\ast ,
{\bf 4}\\ (3,8),{\bf 8}
\end{array}
&
\begin{array}{c}(2,10)^\ast ,{\bf 4}\\(3,9),{\bf 8}\end{array}
&{\underline{(3,10)}}^\ast ,{\bf 8}\\
 \hline
 \end{array}\nonumber
\end{eqnarray}
\begin{eqnarray}
&&
\end{eqnarray}
We have already recalled in section {\bf 4} that for the $(t,s)$
space-times allowing an octonionic description, due to octonionic
non-associative identities, the algebra generated by the
commutators between Gamma matrices is not the $SO(t,s)$ Lorentz
algebra, but its $G_2$ coset $SO(t,s)/G_2$ \cite{hl}. We present
here a generalization of this result consisting of a list of
higher-rank antisymmetric octonionic tensorial identities. It is
worth mentioning that these identities have striking applications
which we shall discuss in the next section.\par The identities
under consideration are applicable to the space-time signatures
which, for a given total dimension $D$, admit spinors of minimal
octonionic dimensionality (up to $D=13$, these are the signatures
denoted with a ``$^\ast$" in the table above). The generalization
of this construction to dimensions $D> 14$ is straightforward.
Here however, both for simplicity and for physical relevance, we
limit ourselves to discuss such identities up to $D=13$, namely
for the following spacetimes
\begin{eqnarray}\begin{array}{|c|c|}\hline
D=7 & (0,7), ~(7,0)\\ \hline D=8& (0,8), (8,0)\\ \hline D=9 &
(0,9), (9,0), (1,8), (8,1) \\ \hline D=10& (1,9), (9,1) \\ \hline
D=11 &(1,10), (10,1),(2,9),(9,2)
\\ \hline D=12& (2,10), (10,2) \\ \hline D=13& (3,10),(10,3),(2,11), (11,2)\\ \hline

\end{array}
\end{eqnarray}
Please notice that in $D=8, 10, 12$ dimensions we are dealing with
fundamental Weyl spinors.\par It is worth mentioning that the
above table has been complemented, for $D=13$, with the non-maximal
octonionic Clifford algebras $(10,3)$, $(2,11)$, arising from the maximal ones
at the level
$D= 15$. \par In the above cases for $D=7,8$ the fundamental
spinors are $1$ (octonionic)-dimensional, $2$-dimensional for
$D=9,10$, four-dimensional  for $D=11,12$ and finally
$8$-dimensional for $D=13$. The total number of octonionic
hermitian ${\cal H}$ (antihermitian ${\cal A}$) components in a
squared matrix of ${\bf d_\Psi}$ size is therefore given by
{{\begin{eqnarray}&
\begin{tabular}{|c|c|c|}\hline
&${\cal H}$&${\cal A}$\\ \hline $D=7,8$&$1$&$7$\\ \hline $D=9,10$
&$10$&$22$\\ \hline $D=11,12$ &$52$&$76$\\ \hline $D=13$
&$232$&$280$\\ \hline
\end{tabular}&\nonumber\end{eqnarray}}}
\begin{eqnarray}
&&
\end{eqnarray}
\par
The antisymmetric product of $k>2$ octonionic $\Gamma$-matrices
must be consistently specified to take into account the
non-associativity of the octonions. As we soon motivate,  the
correct prescription is taking the antisymmetrized product of $k$
octonionic matrices $\Gamma_i$ ($i=1,2,\dots, k$) to be given by
\begin{eqnarray}
\relax [\Gamma_{1}\cdot \Gamma_{2}\cdot \dots \cdot \Gamma_k]
&\equiv& \frac{1}{k!}\sum_{perm.} (-1)^{\epsilon_{i_1\dots i_k}}
(\Gamma_{i_1}\cdot \Gamma_{i_2}\dots \cdot \Gamma_{i_k}),
\label{antisym}
\end{eqnarray}
where $(\Gamma_1\cdot \Gamma_2\dots \cdot \Gamma_k)$ denotes the
symmetric product
\begin{eqnarray}
(\Gamma_1\cdot \Gamma_2 \cdot\dots  \cdot \Gamma_k) &\equiv&
\frac{1}{2}(. ((\Gamma_1 \Gamma_2)\Gamma_3\dots )\Gamma_k)
+\frac{1}{2} (\Gamma_1(\Gamma_2(\dots \Gamma_k)).).
\end{eqnarray}
The usefulness of this prescription is due to the fact that the
product \begin{eqnarray}& A\relax [\Gamma_{1}\cdot \Gamma_{2}\cdot
\dots \cdot \Gamma_k],& \label{agammas}
\end{eqnarray}
with $A$ the matrix (product of the time-like Gamma matrices)
introduced in section {\bf 5} has a definite (anti)-hermiticity
property. The different (\ref{agammas}) tensors, for different choices
of the Gamma's, are all hermitian or antihermitian, depending only
on the value of $k$ (not of the $\Gamma$'s themselves).
\par
In the presence of the Weyl spinors, the above (\ref{agammas}) tensors
can be bracketed with the $P_+$ projection operator, see
(\ref{project}), to give
\begin{eqnarray}&
P_+A\relax [\Gamma_{1}\cdot \Gamma_{2}\cdot \dots \cdot
\Gamma_k]P_+.&\label{aweyls}
\end{eqnarray} Once taken into account, from the
algorithmic table (\ref{bigtable}) applied to the octonionic Clifford
algebras, that out of the $D$ Gamma matrices, $7$ are proportional
to the imaginary octonions, while the remaining $D-7$ are purely
real, it is a matter of straightforward computations to check the
number of independent octonionic components both for (\ref{agammas})
(in the $NW$ spacetimes) and for (\ref{aweyls}) (in the Weyl
spacetimes).
\par In odd-dimensions $D$ we get the table, whose columns are
labeled by the antisymmetric tensors rank $k$,
{{\begin{eqnarray}&
\begin{tabular}{|c|c|c|c|c|c|c|c|c|c|c|c|c|c|c|}\hline
&$0$&$1$&$2$&$3$&$4$&$5$&$6$&$7$&$8$&$9$&$10$&$11$&$12$&$13$\\
\hline $D=7$&${\underline {1}}$&$7$&$7$&${\underline
{1}}$&${\underline{ 1}}$&$7$&$7$&${\underline{ 1}}$&&&&&&\\ \hline

$D=9$&${\underline {1}}$&${\underline {9}}$&$22$&$22$&$
{\underline
{10}}$&${\underline {10}}$&$22$&$22$&${\underline {9}}$&${\underline
{1}}$&&&&\\ \hline

$D=11$&$1$&${\underline {11}}$&${\underline
{41}}$&$75$&$76$&${\underline {52}}$&${\underline
{52}}$&$76$&$75$&${\underline {41}}$&${\underline {11}}$&$1$&&\\
\hline

$D=13$&$1$&$13$&${\underline {64}}$&${\underline{
168}}$&$267$&$279$&${\underline{ 232}}$&${\underline{
232}}$&$279$&$267$&${\underline{ 168}}$&${\underline{
64}}$&$13$&1\\ \hline

\end{tabular}&\nonumber\end{eqnarray}}}
\begin{eqnarray}
&&
\end{eqnarray}
The hermitian components are underlined.\par Similarly, in the
even-dimensional Weyl case, we have

{{\begin{eqnarray}&
\begin{tabular}{|c|c|c|c|c|c|c|c|c|c|c|c|c|c|}\hline
&$0$&$1$&$2$&$3$&$4$&$5$&$6$&$7$&$8$&$9$&$10$&$11$&$12$\\ \hline
$D=8$&${\underline {1}}$&$0$&$7$&$0$&${\underline{
1}}+{\underline{1}}$&$0$&$7$&$0$&${\underline{ 1}}$&&&&\\ \hline

$D=10$&$0$&${\underline {10}}$&$0$&$22$&$0$&${\underline
{10}}+{\underline{10}}$&$0$&$22$&$0$&${\underline {10}}$&$0$&&\\
\hline

$D=12$&$1$&$0$&${\underline {52}}$&$0$&$75$&$0$&${\underline
{52}}+{\underline {52}}$&$0$&$75$&$0$&${\underline
{52}}$&$0$&$1$\\ \hline

\end{tabular}&\nonumber\end{eqnarray}}}
\begin{eqnarray}
&&
\end{eqnarray}
The above tables show the existence of identities relating
higher-rank antisymmetric octonionic tensors. Let us discuss a
specific example, which is perhaps the most physically relevant.
In $D=11$ dimensions the $52$ independent components of an
octonionic hermitian $(4\times 4)$ matrix can be expressed either
as a rank-$5$ antisymmetric tensors (simbolically denoted as
``$M5$"), or as the combination of the $11$ rank-$1$ ($M1$) and
the $41$ rank-$2$ ($M2$) tensors. The relation between $M1+M2$ and
$M5$ can be made explicit as follows. The $11$ vectorial indices
$\mu$ are split into $4$ real indices, labeled by $a,b,c,\ldots$
and $7$ octonionic indices labeled by $i,j,k,\ldots$. We get, on
one side, {{\begin{eqnarray}&
\begin{tabular}{cc}

$4$& $M1_a$\\

$7$&$M1_i$\\

$6$&$M1_{[ab]}$\\

$4\times 7= 28$&$M2_{[ai]}$\\

$7$& $ M2_{[ij]}\equiv M2_{i}$

\end{tabular}&\nonumber\end{eqnarray}}}

while, on the other side, {{\begin{eqnarray}&
\begin{tabular}{cc}

$7$& $M5_{[abcdi]} \equiv M5_i$\\

$4\times 7=28$&$M5_{[abcij]}\equiv M5_{[ai]}$\\

$6$&$M5_{[abijk]}\equiv M5_{[ab]}$\\

$4$&$M5_{[aijkl]}\equiv {M5}_a$\\

$7$& $ M5_{[ijklm]}\equiv {\widetilde M5}_{i}$

\end{tabular}&\nonumber\end{eqnarray}}}
which shows the equivalence of the two sectors, as far as the
tensorial properties are concerned. Please notice that the correct
total number of $52$ independent components is recovered
\begin{eqnarray}
52 &=& 2\times 7 +28+6+4.
\end{eqnarray}
The octonionic equivalence of different antisymmetric tensors can
be symbolically expressed, in odd space-time dimensions, through
{{\begin{eqnarray}&
\begin{tabular}{|c|c|}\hline

$D=7$& $M0\equiv M3$\\ \hline

$D=9$&$M0+M1\equiv M4$\\ \hline

$D=11$&$M1+M2\equiv M5$\\ \hline

$D=13$&$M2+M3\equiv M6$\\ \hline

$D=15$&$M3+M4\equiv M0+M7$\\ \hline

\end{tabular}&\label{tablem}\end{eqnarray}}}

We end up this section by commenting that, for non-minimal spinors, the dependance
on the rank $k$ alone of the hermitian or antihermitian character
of (\ref{agammas}) and (\ref{aweyls}) is not mantained. To be explicit, in $D=8$ space-time dimension,
the spinors associated to the $(1,7)$ signature are non-minimal (the number of
their components is twice the number of components for fundamental
$(8,0)$ and $(0,8)$ spinors). The $(1,7)$ Clifford algebra is obtained from the $(1,8)$ Clifford algebra after deleting a spacelike matrix $\Gamma_S$. For what concerns tensors,
e.g. two sets of vectors are found, the ones obtained from $\Gamma_{\mu}$ ($\mu$ a vector
index in $(1,7)$) are hermitian, while the ones obtained from the commutators
$\relax [\Gamma_{\mu},\Gamma_s]$ are antihermitian.
\section{An application of the octonionic spinors. The octonionic
$M$-algebra and the generalized supersymmetries.}

We shortly review here what is perhaps the most promising
application of the octonionic spinors, i.e. their connection with the
octonionic $M$-algebra (and superconformal $M$ algebra, see
\cite{{lt1},{lt2}}), a specific example of a generalized
octonionic supersymmetry. The identities for antisymmetric
octonionic tensors play in this case a special role.\par The
generalized space-time supersymmetries are the ones going beyond
the standard H{\L}S scheme \cite{hls}. This implies that the
bosonic sector of the Poincar\'e or conformal superalgebra no
longer can be expressed as the tensor product structure
$B_{geom}\oplus B_{int}$, where $B_{geom}$ describes space-time
Poincar\'e or conformal algebras and the remaining generators
spanning $B_{int}$ are Lorentz-scalars.\par In the particular case
of the Minkowskian $D=11$ dimensions, where the $M$-theory should
be found, the following construction is allowed. The spinors are
real and have $32$ components.\par As recalled in section {\bf 2},
by taking the anticommutator of two such spinors the most general
expected result consists of a $32\times 32$ symmetric matrix with
$32+\frac{32\cdot31}{2}=528$ components. On the other hand, the
standard supertranslation algebra underlining the maximal
supergravity contains only the $11$ bosonic Poincar\'e generators
and by no means the r.h.s. saturates the total number of $528$.
The extra generators that should be expected in the right hand
side are obtained by taking the totally antisymmetrized product of
$k$ Gamma matrices (the total number of such objects is given by
the Newton binomial ${\textstyle{\left(\begin{array}{c}
  D \\
  k
\end{array}\right)}}$).
Imposing on the most general $32\times 32$ matrix the further
requirement of being symmetric, the total number of $528$ is
obtained by summing the $k=1$, $k=2$ and $k=5$ sectors, so that
$528=11+55+462$. The most general supersymmetry algebra in $D=11$
can therefore be presented as
\begin{equation}
\{Q_a,Q_b\}= (A\Gamma_\mu )_{ab}P^\mu +(A\Gamma_{[\mu\nu]})_{ab}
Z^{[\mu\nu]} +
(A\Gamma_{[\mu_1\dots\mu_5]})_{ab}Z^{[\mu_1\dots\mu_5]}
\label{Malg}
\end{equation}
(where $A$ is the real matrix introduced in section {\bf 5}).\par
$Z^{[\mu\nu]}$ and $Z^{[\mu_1\dots\mu_5]}$ are tensorial central
charges, of rank $2$ and $5$ respectively. These two extra central
terms on the right hand side correspond to extended objects
\cite{{bst},{dk}}, the $p$-branes. The algebra (\ref{Malg}) is
called the $M$-algebra. It provides the generalization of the
ordinary supersymmetry algebra recovered by setting $Z^{[\mu\nu]}
\equiv Z^{[\mu_1\dots\mu_5]}\equiv 0$.\par On the other hand, in
the same $11$-dimensional Minkowskian spacetime, we can impose, as
we have seen, an octonionic structure, with fundamental spinors
assumed to be $4$-component octonionic valued. The generalized
supersymmetry algebra (\ref{gensusyalg}) admits on the r.h.s. a hermitian
$4\times 4$ octonionic-valued matrix with up to $52$ independent
components. They can be expressed, from the previous section
results, either as the $11+41$
bosonic generators entering
\begin{equation}\label{eq1}
  {\cal{Z}}_{ab} = P^\mu (C\Gamma^{}_\mu)_{ab} +
   Z^{\mu\nu}_{\bf{O}} (C\Gamma^{}_{\mu\nu})_{ab}
   ,
\end{equation}
or as the $52$ bosonic generators entering
\begin{equation}\label{eq2}
  {\cal{Z}}_{ab} =
    Z^{[\mu_1\ldots \mu_5]}_{\bf{O}}
    (C\Gamma^{}_{\mu_1 \ldots
    \mu_5})_{ab}\, .
\end{equation}
Differently from the real case,
the sectors specified by (\ref{eq1}) and (\ref{eq2}) are not
independent\cite{lt1}, leading to an unexpected and far from trivial
new structure in the octonionic $M$-algebra.\par
The octonionic results contained in the present paper should be
regarded as the necessary background towards a classification of
the octonionic generalized supersymmetries which is at present
still missing. For what concerns the $1D$ octonionic supersymmetries \cite{top}
applied to octonionic quantum mechanics, a classification is now available
\cite{crt}.

\section{Conclusions.}

In this paper we made a systematic investigation of real,
quaternionic and octonionic-valued Clifford algebras and spinors,
presenting their classification, as well as constructive formulas
to iteratively produce them. Tables have been given with the most
general free dynamics satisfied by real, quaternionic and
octonionic spinors in each space-time which supports them. All
kinetic and massive terms have been listed.\par For what concerns
the octonionic case, by far the most intriguing due to the
non-associativity, we further presented the systematic
construction and derived a series of tables expressing the
identities among higher rank antisymmetric octonionic tensors. We
motivated this line of research with the attempt at classifying
the generalized octonionic supersymmetries. A first example,
hopefully physically relevant, consists of the octonionic
$M$-algebra, with its striking properties induced by the mentioned
identities.\par For what concerns the quaternionic spinors, they
also can appear in connection with generalized supersymmetries.
One can read, e.g., from the results here presented, that in the
Euclidean $D=11$ dimensions quaternionic-valued spinors are
allowed. It looks promising to employ them to construct a
quaternionic Euclidean version of the $M$ algebra (we are in fact
planning to address this problem in the future).\par Coming back
to the octonionic spinors, we mention a further list of topics
where they can possibly find application. At the end of section
{\bf 3} we pointed out that the octonionic realization of the
$7$-dimensional Euclidean Clifford algebra is related with the
geometry of the seven sphere $S^7$. A question, which deserves
being investigated, can be raised. Is the octonionic description
of the $M$-theory somehow related to the particular
compactification of the $11$-dimensional $M$-theory down to
$AdS_4\times S^7$? This compactification corresponds to a natural
solution for the $11$-dimensional supergravity \cite{dp}. It would
be interesting to check whether the tensorial identities found in
the octonionic construction find a counterpart also in the
$AdS_4\times S^7$ special compactification geometry. On the other
hand, one should try to understand the physical implications of
the octonionic $M$-algebra also from a purely algebraic point of
view. Being expressed by a $4$-dimensional octonionic matrix, it
is outside a Jordan algebra scheme \cite{gpr}. This raises the
question of its quantum-mechanical consistency, which implies
understanding whether, and to which extent, is it possible to
adapt the prescription of \cite{gpr} to the present situation.\par
It is worth mentioning a different dynamical system \cite{smo},
which can be called a ``Jordan Matrix Chern-Simon theory",
proposed as a unique model, being associated with the exceptional
Jordan algebra $J_3({\bf O})$ of $3\times 3$ hermitian octonionic
matrices. In this context it seems relevant addressing, for
octonionic fields, the status of the spin-statistic theorem, in
order to carefully revise it. Throughout this paper we have
assumed the octonionic spinors being Grassmann, anticommuting
fields.  However, it cannot be a priori excluded that in the
octonionic case this assumption could be relaxed.\par We finally
mention that the octonions can be held responsible for the
existence of a bunch of exceptional structures in Mathematics. As
an example the $5$ exceptional Lie algebras can all be produced
from the octonions via the Tits' construction \cite{bs}. A lot of
activity is currently devoted to explore the relevance for Physics
of these exceptional structures \cite{ram}, see also \cite{ohw}.
The octonions seem the right tool to investigate such connections,
see e.g. \cite{boy}. The recognized importance of this line of
research strongly motivated us to systematically present here the
fundamental properties of octonionic fields and spinors, as well
as their non-trivial relations, as the ones discussed in section
{\bf 9}.
\\$~$
\par
 {\large{\bf Acknowledgments.}} ~\\~\par

We acknowledge very useful discussions with J. Lukierski.


\begin{thebibliography}{9}
\bibitem{bv} E. Bergshoeff and A. Van Proeyen, Class. Quant.
Grav. {\bf 17}, 3277 (2000).
\bibitem{gsw} M.B. Green, J.H. Schwarz and E. Witten, ``Superstring Theory"
($2$ Volumes), Cambr. Univ. Press (1987).
\bibitem{abs} M.F. Atiyah, R. Bott and A. Shapiro, Topology
(Suppl. 1) {\bf 3}, 3 (1964).
\bibitem{kt} T. Kugo and P. Townsend, Nucl. Phys. {\bf B 221}, 357
(1983).
\bibitem{oku1} S.Okubo, J. Math. Phys. {\bf 32}, 1657 (1991).
\bibitem{fer} R. D'Auria, S. Ferrara and S. Vaula, Class. Quant.
Grav. {\bf 18}, 3181 (2001); JHEP 0010 (2000), 013; S. Ferrara,
M.A. Lledo, hep-th/0112177.
\bibitem{hls} R. Haag, J. \L opusza\'{n}ski and M. Sohnius, Nucl.Phys.
{\bf B 88}, 257 (1975).
\bibitem{lt1} J. Lukierski and F. Toppan, Phys. Lett. {\bf B 539},
266 (2002).
\bibitem{lt2} J. Lukierski and F. Toppan, ``Octonionic $M$-theory
and $D=11$ Generalized Conformal and Superconformal Algebras", hep-th/0212201.
\bibitem{jor} P. Jordan, Nachr. Ges. Wiss. G\"{o}tt. Math. Phys. 569
(1932), {\em ibid.} 209 (1933); Zeits. Phys. {\bf 80}, 285 (1933); P. Jordan,
J. von Neumann and E.P. Wigner, Ann. Math. {\bf 35}, 29 (1934).
\bibitem{fm} D.B. Fairlie and A.C. Manogue, Phys. Rev. {\bf D 34},
1832 (1986).
\bibitem{cs} K.W. Chung and A. Sudbery, Phys. Lett. {\bf B 198},
161 (1987).
\bibitem{oku2} S. Okubo, J. Math. Phys.{\bf 32}, 1669 (1991).
\bibitem{hl} Z. Hasiewicz and J. Lukierski, Phys. Rev. {\bf 145
B}, 165 (1984).
\bibitem{cp}  M. Cederwall and C. Preitschopf, Comm. Math. Phys. {\bf
167}, 373 (1995).
\bibitem{dt} M.A. De Andrade and F. Toppan, Mod. Phys. Lett. {\bf A 14},
1797 (1999).
\bibitem{gk} M. G\"{u}naydin and S.V. Ketov, Nucl. Phys. {\bf B
467}, 215 (1996).
\bibitem{bst} E. Bergshoeff, E. Sezgin, P, Townsend, Ann. Phys.
{\bf 18}, 330 (1987).
\bibitem{dk} M.J. Duff. R.R. Khuri, J.X. Lu, Phys. Rep. {\bf
259}, 213 (1995).
\bibitem{top} F. Toppan, Nucl. Phys. B (Proc.
Suppl.) {\bf 102\&103}, 270 (2001).
\bibitem{crt} H.L. Carrion, M. Rojas and F. Toppan,
``Octonionic Realizations of $1$-dimensional Extended Supersymmetries. A
Classification", hep-th/0212030.
\bibitem{dp} M.J. Duff and C.N. Pope, ``Kaluza-Klein supergravity
and the seven-sphere" in Supersymmetry and Supergravity 82,
Trieste proceedings; M.J. Duff, B.E.W. Nilsson and C.N. Pope,
Phys. Rep. {\bf 130}, 1 (1986).
\bibitem{gpr} M. G\"{u}naydin, C. Piron and H. Ruegg, Comm. Math.
Phys. {\bf 61}, 69 (1978).
\bibitem{smo} L. Smolin, hep-th/0104050.
\bibitem{bs} C.A. Barton and A. Sudbery, math.RA/0203010.
\bibitem{ram} P. Ramond, ``Algebraic Dreams", hep-th/0112261.
\bibitem{ohw} Y. Ohwashi, Prog. Theor. Phys. {\bf 108}, 755 (2002).
\bibitem{boy} L.J. Boya, ``Octonions and $M$-theory", hep-th/0301037.

\end{thebibliography}
\end{document}